%% file: main.tex
\documentclass[sigconf]{acmart}

\makeatletter
\def\@ACM@checkaffil{%
  \if@ACM@instpresent\else\ClassWarningNoLine{\@classname}{No institution present}\fi
  \if@ACM@citypresent\else\ClassWarningNoLine{\@classname}{No city present}\fi
  \if@ACM@countrypresent\else\ClassWarningNoLine{\@classname}{No country present}\fi}
\makeatother

\AtBeginDocument{%
  }
    
\copyrightyear{2025}
\acmYear{2025}
\setcopyright{cc}
\setcctype{by-nc}
\acmConference[PASC '25]{Platform for Advanced Scientific Computing Conference}{June 16--18, 2025}{Brugg-Windisch, Switzerland}
\acmBooktitle{Platform for Advanced Scientific Computing Conference (PASC '25), June 16--18, 2025, Brugg-Windisch, Switzerland}
\acmDOI{10.1145/3732775.3733586}
\acmISBN{979-8-4007-1886-1/2025/06}

\usepackage{amsmath,amsfonts}
\usepackage{algorithmic}
\usepackage{graphicx}
\usepackage{xcolor}
\usepackage{tabularx}
\usepackage[detect-all,group-separator={,}]{siunitx}
\usepackage{url}
\usepackage{enumitem}
\usepackage{textcomp}
\usepackage{bibentry}
\usepackage{natbib}
\usepackage{color}
\usepackage{xspace}
\usepackage{bigstrut}
\usepackage{multirow}
\usepackage{subcaption}
\usepackage{balance}
\usepackage{makecell}
\usepackage{hyperref}
\usepackage{float}
\usepackage{caption}

\newcommand{\aglimmer}{HiPerRAG}
\newcommand{\OREO}{Oreo}

\newcommand{\SFRmistral}{SFR-Embedding-Mistral}

\begin{document}


\title{HiPerRAG: \underline{Hi}gh-\underline{Per}formance \underline{R}etrieval \underline{A}ugmented \underline{G}eneration for Scientific Insights}


\author{
Ozan Gokdemir$^{1,2\dagger^{*}}$, 
Carlo Siebenschuh$^{1,2\dagger}$,  
Alexander Brace$^{1,2\dagger}$,
Azton Wells$^{1\dagger}$,
Brian Hsu$^{1,2\dagger}$,
Kyle Hippe$^{1,2}$,
Priyanka V. Setty$^{1,2}$,
Aswathy Ajith$^{2}$,
J. Gregory Pauloski$^{2}$,
Varuni Sastry$^{1}$,
Sam Foreman$^{1}$,
Huihuo Zheng$^{1}$,
Heng Ma$^{1}$,
Bharat Kale$^{1}$,
Nicholas Chia$^{1}$,
Thomas Gibbs$^{3}$,
Michael E. Papka$^{1,4}$,
Thomas Brettin$^{1}$,
Francis J. Alexander$^{1}$,
Anima Anandkumar$^{5}$,
Ian Foster$^{1,2}$,
Rick Stevens$^{1,2}$,
Venkatram Vishwanath$^{1}$,
Arvind Ramanathan$^{1,2}$,
}

\affiliation{
 $^{1}$Argonne National Laboratory, Lemont, Illinois, USA
 $^{2}$The University of Chicago, Chicago, Illinois, USA \and
 $^{3}$NVIDIA Inc., Santa Clara, California, USA 
 $^{4}$University of Illinois Chicago, Chicago, Illinois, USA \and
 $^{5}$California Institute of Technology, Pasadena, California, USA
}

\affiliation{$\dagger$Joint first authors $^{*}$Corresponding authors: ramanathana@anl.gov, stevens@anl.gov, venkat@anl.gov}

\renewcommand{\shortauthors}{Gokdemir, et al.}

\begin{abstract}
The volume of scientific literature is growing exponentially, leading to underutilized discoveries, duplicated efforts, and limited cross-disciplinary collaboration. Retrieval-Augmented Generation (RAG) offers a way to assist scientists by improving the factuality of Large Language Models (LLMs) in processing this influx of information. However, scaling RAG to handle millions of articles introduces significant challenges, including the high computational costs associated with parsing documents and embedding scientific knowledge, as well as the algorithmic complexity of aligning these representations with the nuanced semantics of scientific content. To address these issues, we introduce \aglimmer, a RAG workflow powered by high performance computing (HPC) to index and retrieve knowledge from more than 3.6 million scientific articles. At its core are \OREO, a high-throughput model for multimodal document parsing, and ColTrast, a query-aware encoder fine-tuning algorithm that enhances retrieval accuracy by using contrastive learning and late-interaction techniques. \aglimmer~delivers robust performance on existing scientific question answering (Q/A) benchmarks and two new benchmarks introduced in this work, achieving ~90\% accuracy on SciQ and 76\% on PubMedQA---outperforming both domain-specific models like PubMedGPT and commercial LLMs such as GPT-4. Scaling to thousands of GPUs on the Polaris, Sunspot, and Frontier supercomputers, \aglimmer~delivers million document-scale RAG workflows for unifying scientific knowledge and fostering interdisciplinary innovation.
\end{abstract}



\settopmatter{printacmref=true}
\setcopyright{cc}
\renewcommand\footnotetextcopyrightpermission[1]{}
\pagestyle{plain}

\maketitle

\section{Introduction}

Over the past century, the volume of scientific publications has increased rapidly. Today, this trend continues at an unprecedented pace. For example, the National Science Foundation reported a 50-fold increase in open-access scientific publications between 2013 and 2022 \cite{nsb2024}. In the biomedical field alone, PubMed processed approximately 1.69 million articles last year, averaging more than three articles per minute \cite{landhuis2016}. Despite this vast output, researchers' capacity to keep up is limited, estimated at 22 articles per month \cite{vannoorden2024}. This overload of information has significant consequences: more than half of the 200 million scholarly papers available today have never been cited \cite{SinghChawla2022}. Meanwhile, the top 1\% of most cited scientists cumulatively receive 21\% of all citations \cite{nielsen2021}. This imbalance leads to a lack of diversity in academic ideas, redundant research efforts, and hindered scientific progress \cite{chu2021slowed}. Addressing these challenges requires innovative approaches to ingesting and disseminating scientific knowledge more effectively. In particular, machine learning poses a promising path towards helping researchers navigate this rapid influx of information.

The advent of large language models (LLMs) has provided researchers with performant tools for processing vast scientific corpora. LLMs have been explored to help with knowledge-intensive tasks, such as question answering \cite{Akinseloyin2024, Auer2023}, information retrieval \cite{Akinseloyin2024}, document summarization \cite{chang_summary, godbole_summary}, and generating novel hypotheses \cite{Banker2024}. Despite their utility, a crucial complication emerges when LLMs are applied to science, mainly, the need for any generated output to be grounded in factuality. Unfortunately, LLMs are prone to hallucinate plausible yet factually incorrect information, and science-based LLMs showcase limitations that prevent their wide-scale adoption in the scientific enterprise \cite{taylor2022galactica, goswami2023switchprompt}. In particular, enforcing factuality and reducing hallucination is an ongoing challenge involving multiple avenues of research \cite{gao2023retrieval,chuang2023dola,asai2023self}.

The introduction of Retrieval-Augmented Generation (RAG) \cite{rag} aims to mitigate hallucinations of LLMs through the integration of neural information retrieval with LLM-based content generation. This approach leverages the ability of LLMs to represent any chunk of text---a word, a sentence, a paragraph, or a whole document---as a fixed-size embedding vector that encodes semantic relationships akin to meaning. The distance between these embedding vectors is used as a measure of relevance between two texts such as a question and a scientific passage answering it. Quantifying relevance in this manner improves the factuality of LLM outputs by dynamically equipping LLMs with contextually relevant passages during the generation process \cite{karpukhin2020densepassageretrievalopendomain}.

\begin{figure}[ht]
    \centering
    \includegraphics[width=8.7cm]{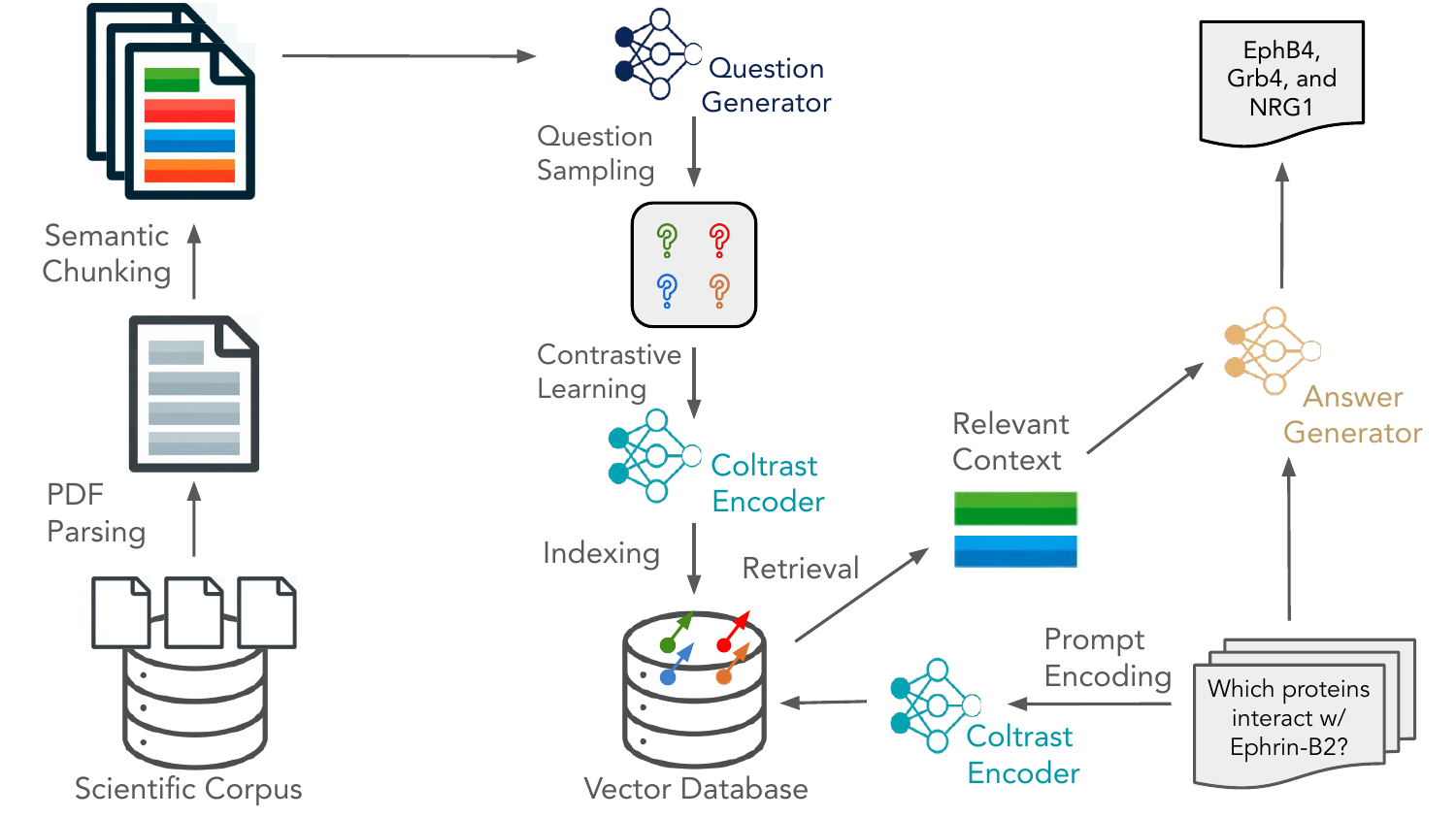}
    \caption{\aglimmer~ Workflow. A graphical overview of our implementation of RAG for scientific literature. We implement a novel PDF parsing approach, namely optical recognition with eclectic output (Oreo) that takes into account intrinsic formatting of multi-layout scientific manuscripts. We then design a query-aware encoder finetuned on scientific literature which enables to semantically organize the data into relevant chunks that can be used to retrieve the most relevant data on a per-query basis. Both these innovations enable us to achieve both computational efficiency and retrieval performance at scale.}
    \label{fig:auroraglimmer} 
\end{figure}

Despite the widespread adoption of RAG, it faces three significant technical challenges that hinder its ability to scale to millions of documents. These include: 
\begin{enumerate}[leftmargin=*]
    \item \textbf{Parsing meaningful and coherent text from scientific documents:} This challenge emerges from the diverse organization semantics of how papers are published, including dense arrangements of tables, figures, and equations in a portable document format (PDF) file. This process requires intricate subtasks that range from image normalization to character recognition, making it a complex endeavor.
    \item \textbf{Optimizing the retrieval accuracy using LLMs:} LLMs need to be provided with the most relevant information---and only that---to answer the specific query at hand. This crucial task depends on an encoder’s ability to produce vector representations of both user queries and the document corpus. General purpose encoders perform suboptimally when encoding scientific content since such content constitutes only a small portion of mainstream LLM training datasets \cite{dolma, gao2020pile800gbdatasetdiverse}.
    \item \textbf{Evaluating RAG systems on scientific literature:} Scientific literature carries unique features including domain-specific vocabulary, empirical evidence, and citations, which are generally hard to assess. Hence this imposes certain restrictions on how RAG can be evaluated on scientific literature, where fetching both the right context and the appropriate related content can be difficult. 
\end{enumerate}

\paragraph{\textbf{Contributions}:} Given these limitations, our work, \aglimmer\footnote{https://github.com/ramanathanlab/distllm} embodies innovations on several components of the RAG framework for scientific literature, including document parsing, encoding scientific information, and science-specific evaluation benchmarks to overcome each of the aforementioned challenges (see \autoref{fig:auroraglimmer}). Our contributions are summarized as follows: 


\begin{itemize}[leftmargin=*]
\item We develop neural Optical Recognition with Eclectic Output (Oreo), a layout-aware multimodal scientific document parser for high-quality content extraction from PDFs. \OREO~ outperforms state-of-the-art parsers in terms of throughput while maintaining a commensurate accuracy. \autoref{subsec:document-parsing} and \autoref{subsec:oreo_eval} discuss \OREO's design, implementation, and evaluation.  

\item We design a query-aware encoder tailored for high accuracy on scientific content that is both state-of-the-art and performs at scale. To accomplish this we develop a novel finetuning algorithm for scientific text encoding, called ColTrast, which combines contrastive and late-interaction methods to capture the benefits of both to custom-tailor passage retrieval to user queries. \autoref{sec:coltrast_design} presents ColTrast's design and implementation and \autoref{subsec:coltrast_eval} presents its evaluation results.


\item We introduce two new biomedical Q/A benchmarks for RAG containing question-answering pairs for protein interaction and function predictions, namely the ProteinInteractionQA (\num{7591} Q/A pairs) and ProteinFunctionQA (\num{17646} Q/A pairs). \autoref{sec:qa-datasets} discusses their curation in detail. We also introduce a synthetic dataset, BioSynthQPs, consisting of \num{1500} domain-specific scientific passages, to evaluate the retrieval accuracy of encoders. \autoref{sec:qa-datasets} describes the process of its creation.

\item We leverage the Parsl~\cite{babuji19parsl} framework to distribute document parsing, encoding, and retrieval components of a RAG workflow to an arbitrary number of nodes on diverse HPC systems. \autoref{sec:scaling_results} contains scaling experiments on the Polaris, Sunspot, and Frontier supercomputers.

\end{itemize}

The remainder of this paper is organized as follows. \autoref{sec:backgound} surveys the current state of the art in document parsing, neural information retrieval, and retrieval-augmented generation. \autoref{sec:design} presents the detailed design and implementation of the individual components of \aglimmer. \autoref{sec:eval} evaluates the \aglimmer~system, detailing the performance of the OREO parser and ColTrast encoder individually, and the overall efficacy of the system on scientific Q/A tasks. It also introduces the Q/A and retrieval datasets used for testing. \autoref{sec:scaling_results} discusses the performance attributes of \aglimmer~and presents the scaling experiments on three supercomputers. \autoref{conclusions} concludes the paper with a discussion of current limitations, future research directions, and implications.


\section{Background}\label{sec:backgound}

We survey related work on document parsing, neural information retrieval, metric learning, and retrieval-augmented generation.

\subsection{Document Parsing}
The portable document format (PDF) has become the primary mode of scientific communication \cite{Ahmed2017}. Although PDFs can store diverse content including figures, tables, equations, and references, they are not machine-readable. \cite{pdf2016}. Therefore, the content of a PDF must be converted into structured outputs, such as raw text, tabular data, and figures. We refer to this task as document parsing. 

Extraction-based parsers such as PyPDF \cite{pypdf}, PyMuPDF \cite{pymupdf}, and PDFMiner \cite{pdfminer} leverage the internal structure of PDF files to obtain text and metadata. However, while these methods achieve high throughput, they rely on an embedded text layer that is generally only present in born-digital PDFs, i.e., they do not apply to scanned PDFs. Furthermore, their accuracy is sensitive to formatting errors within PDF files, resulting in a high file-level failure rate and low overall accuracy. Finally, extraction-based parsing is mostly restricted to raw text, neglecting tables and figures. 

Optical Character Recognition (OCR) is a more advanced document parsing strategy that generates machine-readable content directly from the image renderings of the PDF pages. Traditional OCR approaches involve several deterministic steps such as image pre-processing (including noise reduction and normalization), segmentation (to separate characters or words), feature extraction (to identify distinctive character features), and eventually culminating in character recognition and extracted text. With the advent of deep learning in computer vision, the use of neural networks has streamlined this process either by replacing portions of the OCR pipeline \cite{xu2020layoutlm} or by training an end-to-end image-to-text decoder \cite{kim2022ocr,blecher2023nougat}. The leading edge in document parsing is Neural Optical Understanding for Academic Documents (Nougat) \cite{blecher2023nougat} that uses a Vision Transformer (ViT) to embed a document page image and a text decoder to generate text from it. Nougat employs a hierarchical ViT to process large-image renderings of PDF pages, essentially encapsulating both low-resolution OCR tasks such as layout identification and high-resolution ones such as character recognition in one end-to-end network architecture. Since hierarchical ViTs learn to decode the contents of a page directly from its image, they require GPU-accelerated training on large high-quality datasets. This requirement introduces scalability challenges. In addition, despite being ViT-based, these approaches are currently unable to process data modalities other than text, such as figures. 

To achieve streamlined ViT-based document parsing, recent methods such as Marker \cite{marker} explored a hybrid of layout-based and OCR-based parsing. Marker first performs layout detection to segment the PDF into text, image, table. These segments are then parsed using the Texify \cite{texify} pre-trained ViT to extract machine-readable content. This model claims to realize an approximately 10-fold improvement over Nougat \cite{blecher2023nougat}, which we could not confirm in our experiments. 

We propose Optical Recognition with Eclectic Output (Oreo), which substantially accelerates document parsing while maintaining comparable accuracy to that of leading-edge parsers.
Oreo achieves this by conducting PDF parsing in two stages. First, it leverages a convolutional neural network architecture (YOLO) for layout detection \cite{redmon2016you}. Second, it applies Texify to regions predicted to be text to retrieve textual content. \OREO~ achieves a $4.5 \times$ speedup and a $94.6 \times$ improvement in FLOP utilization over current state-of-the-art (SOTA) methods, including Marker. It supports multimodal parsing for text, figures, and tables, and maximizes GPU utilization to deliver the highest throughput with comparable accuracy to other vision-based parsers. This efficiency highlights \OREO's capability for targeted content extraction and superior hardware resource leverage.

\subsection{Neural Information Retrieval and Metric Learning}

Neural information retrieval uses compact vector representations of data, known as embeddings, that capture meaningful characteristics
\cite{karpukhin2020densepassageretrievalopendomain, mikolov2013efficientestimationwordrepresentations}. The distance between the embeddings of a query and a document, measured by cosine similarity or Euclidean distance, quantifies their relevance to each other. This measurement allows retrieval systems to efficiently identify and retrieve documents that are most relevant to a specific query by comparing these distance metrics. Since retrieval results depend directly on the embedding vectors, training on a broad scientific text corpus is a prerequisite for an encoder LLM to offer embeddings suitable for accurate retrieval on scientific content.

While Transformer-based LLMs are successful information encoders \cite{lu2021pretrainedtransformersuniversalcomputation}, most general-purpose LLMs are not exposed to enough domain-specific scientific data during training \cite{gao2020pile800gbdatasetdiverse, dolma}. Since academic writing style differs substantially from general data such as news articles, embeddings produced by these LLMs are suboptimal for accurate information retrieval for niche scientific queries. Previous work has enhanced the performance of LLMs on downstream tasks by pre-training encoders from scratch using domain-specific data. This approach has been applied across various fields, including biomedical \cite{pubmedbert, biobert_lee2019}, chemical \cite{Schwaller2019, smiles_transformer, Zhou2023}, and material science \cite{Tshitoyan2019}. While domain-specific training of encoders has proven valuable, incorporating metric learning techniques can further refine retrieval performance by optimizing the encoder to better align query embeddings with the most relevant document embeddings.

Several techniques have been explored for building boutique encoders for domain-specific scientific content. In particular, contrastive learning \cite{contrastive-learning} has improved retrieval accuracy by better estimating the relevance of document passages to queries. Contrastive learning teaches models to distinguish between similar (positive) and dissimilar (negative) pairs, encoding data such that positives are closer and negatives farther apart in feature space. Methods like COSTA \cite{contrastive1} and FILIP \cite{contrastive2} successfully applied this to text and image encoders, respectively.

Late interaction, introduced in ColBERT \cite{khattab2020colbert}, a BERT \cite{kenton2019bert} variant, computes token-level contextualized embeddings by maximizing pairwise similarity between query and text token embeddings. This fine-grained comparison increases semantic alignment and retrieval accuracy, but is computationally expensive, replacing a single tensor operation with $N^2$ tensor operations. ColBERTv2 \cite{santhanam-etal-2022-colbertv2} optimizes this by performing token-wise comparisons only after encoding, enabling offline computation of document embeddings and efficient retrieval using a vector database. While this improves efficiency, it relies on ad-hoc re-ranking and does not scale to end-to-end ranking for large corpora. In this work, we propose the ColTrast algorithm, which combines the late interaction mechanism and contrastive learning to achieve robust retrieval accuracy and scales to thousands of GPUs to cater for millions of documents.

\subsection{Retrieval-Augmented Generation}

Retrieval-Augmented Generation (RAG) is a prominent technique that integrates neural information retrieval methods with large language models (LLMs) to enhance the factuality and relevance of generated outputs. In their seminal work, Lewis et al. \cite{rag} demonstrated the effectiveness of RAG in tasks such as question answering, fact-checking, and knowledge extraction. Introduced at a time when LLMs were already extensively adopted in industry and academia, and their limitations in factuality, relevance, and bias were well-documented, RAG represented a significant advancement in Natural Language Processing. 

A key preliminary step in implementing a RAG application involves chunking and encoding a curated corpus of relevant documents. The resulting chunk embeddings are stored in a vector database such as Faiss \cite{douze2024faiss} or Chroma \cite{chromadb}, each linked to a unique ID that maps to the raw text in a separate database. When a query is provided to an LLM, it is encoded using the same encoder as the document chunks, and the resulting query embedding is used for a semantic search in the vector database to identify the nearest neighbors. Common distance metrics for this search include cosine similarity and Euclidean distance. Finally, the retrieved IDs are used to fetch the raw content of the relevant documents, which are supplied to the LLM alongside the query to generate a response.

RAG has been widely adapted to many scientific domains, including material science \cite{rag_application_1}, biology \cite{wang2024bioragragllmframeworkbiological, Matsumoto2024}, and chemistry \cite{chen2024chemistxlargelanguagemodelempowered}. However, scaling a RAG application up to millions of documents remains an ongoing challenge in both academia and industry. In response, we present \aglimmer, a high-performance computing infrastructure designed to scale every component of RAG to an arbitrary number of compute nodes. With 3.6 million indexed scientific papers, \aglimmer~ represents a step toward integrating approximately 200 million existing scientific papers into a single platform for scientific Q/A, literature review and hypothesis generation.

\section{HiPerRAG Design and Implementation}\label{sec:design}

Despite the continued advancement of Large Language Models (LLMs) in reasoning and creative writing, their limitations in operations like tool usage or information retrieval persist. This has spurred the development of sophisticated software stacks that enhance LLMs with external tools, such as information retrievers, to augment text generation processes and implement Retrieval-Augmented Generation (RAG) applications ~\cite{wu2023autogenenablingnextgenllm, erdogan2024tinyagentfunctioncallingedge, rag}. In response to the scalability challenges in existing RAG frameworks, we introduce \aglimmer---an HPC software stack designed to build scalable, million-document RAG applications for scientific use. \aglimmer~serves not as a standalone model, but as an infrastructure that supports the development of one of the most comprehensive RAG systems on scientific data to date.

To enhance the scalability and accuracy of RAG, our work introduces innovations in PDF parsing and encoder fine-tuning through \OREO~and ColTrast, respectively. In \autoref{subsec:document-parsing} we discuss the implementation details of \OREO. \autoref{sec:coltrast_design} presents the ColTrast algorithm. \autoref{sec:warmstart} will conclude this section with the details of the overall HPC workflow that powers \aglimmer~ and the warmstart optimization we implemented to optimize throughput and GPU utilization. \autoref{fig:auroraglimmer} depicts the high-level RAG workflow implemented by \aglimmer.

\subsection{Document Parsing with \OREO}
\label{subsec:document-parsing}

A system that extracts text from PDFs follows one of two paradigms: the end-to-end approach or the compartmentalized approach. End-to-end document parsing generally relies on the use of multi-modal AI algorithms which, despite making use of GPU acceleration, are challenging to scale to large (>1M) document corpora due to the quadratic scaling of the ViT attention mechanism. However, these methods are not limited to digital-born PDFs with internal layout and metadata. On the other hand, more cost-efficient compartmentalized approach divides the conversion from image to text into several canonical phases which usually rely on the internal layout of the PDF file. Despite having higher throughput for lower cost, these approaches are not applicable to scanned PDF files. The compartmentalized approach generally consists of the following steps: image pre-processing, layout detection, text box extraction, character recognition, and text ordering. 

\begin{figure}[ht]
    \centering
    \includegraphics[width=\columnwidth]{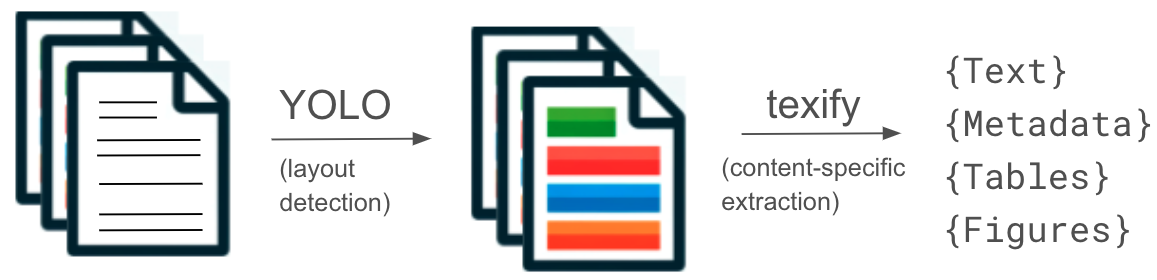}
    \caption{Parsing workflow for scientific PDFs. The neural optical recognition with eclectic output allows content-aware extraction of information from large-scale PDF collections.}
    \label{fig:oreo} 
\end{figure}

We achieve across-the-board improvements in scaling performance in Oreo by combining the strengths of both end-to-end and compartmentalized approaches. PDFs are processed page by page and converted to images. Batches of pages are then parsed in a two-step process.

First, layout detection is performed using convolutional neural networks that identify regions corresponding to text, figures, and table categories. Text regions are further categorized into metadata (e.g., abstract, title, author, publication date) and content (e.g., paragraphs, footnotes). The YOLOv5 architecture used in our experiments provided the best trade-off between throughput, accuracy, and scalability.

Second, Texify is applied to image patches identified as text in the previous step to decode them into plain text. It achieves this by embedding the image using a Vision Transformer (ViT) and decoding the text autoregressively \cite{texify}. The decoded text items are subsequently recombined for each page and merged into the document's full text.


In addition to achieving a $4.5 \times$ speed-up over the current state-of-the-art approach, Marker, \OREO~also surpasses both Nougat and Marker in multimodal capabilities. Specifically, \OREO~can distinguish between 20 types of document assets, compared to Marker's three. Unlike Nougat, which can parse only text from PDFs, \OREO~comprehensively handles a variety of document assets including figures, tables, equations, and code. Furthermore, \OREO~explicitly parses metadata, which enables the storage of citations, author names, and references from scientific documents to facilitate digital archiving. \autoref{fig:oreo} depicts \OREO's overall document parsing workflow.

\subsection{ColTrast: Query-Aware Encoder Finetuning}
\label{sec:coltrast_design}

The discussion of the ColTrast query-aware encoder finetuning algorithm is organized as follows. \autoref{subsubseq:design-choices} presents the design choices we made on the similarity metric, model size, and loss function along with our reasoning for each. \autoref{subsubsec:coltrast_trainset} introduces the ColTrast training dataset and the procedure for curating it. 

\begin{figure}[ht]
    \centering
    \includegraphics[width=\columnwidth]{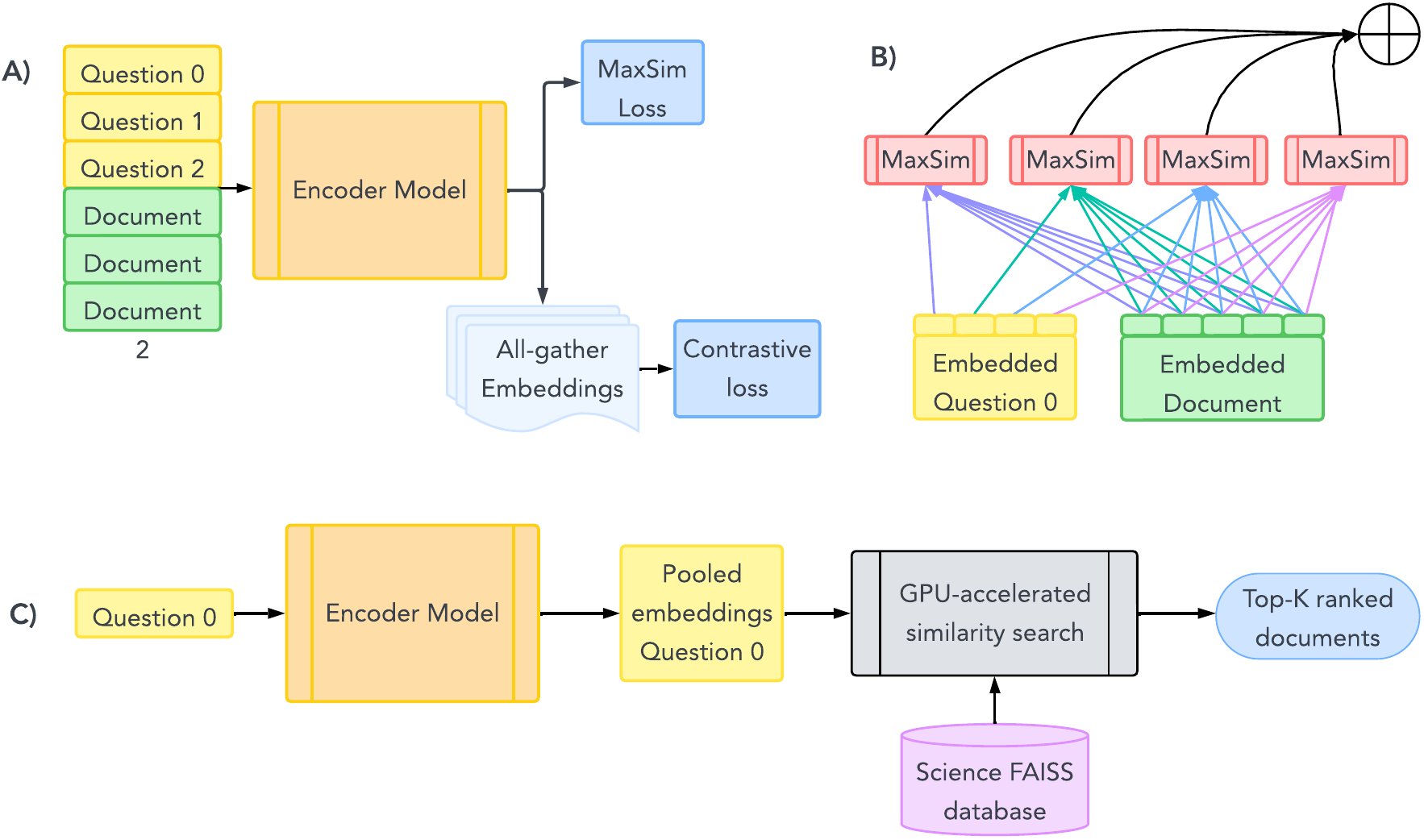}
    \caption{Encoder/retrieval models workflow using the ColTrast loss method. A) Training using ColTrast combines local token-level late-interaction loss with a contrastive loss operating on gathered embeddings. B) Details of late-iteration loss where each token in question/query is compared to each document.  C) Inference time workflow.  Each question is embedded, and then used in similarity search to extract relevant chunks from the scientific database.}
    \label{fig:coltrast-workflow}
\end{figure}

\subsubsection{ColTrast Architecture and Implementation}
\label{subsubseq:design-choices}

Embedding-based retrievers function on the premise that similar vectors indicate similarities in text. Therefore, in many question answering use cases, it is common to embed the query text and identify the most similar document chunk embeddings. The texts corresponding to these embeddings then are added as part of the in-context information the LLM can use to help answer the original question. In other words, the goal is to augment the LLM by providing text containing the information needed to answer the original question. The major determinants of retrieval performance are similarity metric and embedding performance, which is determined by model size, loss function, and training data.

We note that many of these performance determinants are double-edged: cutting back computation and improving scaling often comes at the cost of retrieval accuracy and vice-versa. 
However, we identified several places where we could improve accuracy with no or minimal loss of scaling performance, as described below.

\paragraph{Similarity metric: token-by-token vs.\ pooled embeddings}
Although ColBERT is a performant retriever in terms of retrieval accuracy, its late interaction loss function involves a token-by-token comparison that scales as the product of the numbers of tokens in the document chunk and query. \autoref{fig:coltrast-workflow} illustrates the detrimental effect of this quadratic process for a single comparison in terms of scaling performance. For a large-scale vector store with billions of chunks of text, these scaling barriers become impractical. Ideally, we would like to retain performance and scalability together. Scalabilty in this case depends on choice of similarity metric and we choose to replace the token-by-token similarity metric with the cosine similarity on the pooled embeddings. Normally, this would lead to performance degradation. We describe in the following how we overcame this problem by enhancing the accuracy of the embedding model.

\paragraph{Embedding performance: model size}
For a well-trained model, model size is often a primary determinant of model performance. While BERT-style models such as ColBERT perform well on embedding tasks, especially for their size ($\sim$110M parameters), billion parameter GPT-style models have started to surpass moderate BERT-style models on the embedding leaderboard. At the time of writing, a 7B parameter GPT-style model named \SFRmistral~\cite{Meng_SFR_Mistral} has the best overall performance on embedding tasks, meaning it is the embedding model that most accurately captures semantic meaning appropriately. It is therefore likely that \SFRmistral~ is a good basis for enhancing the performance of our similarity metric that uses pooled embeddings. 

\paragraph{Embedding performance: loss function}
Fine-tuning on relevant tasks can enhance performance. However, increasing the model size by 70$\times$ limits our batch sizes and requires model sharding across GPUs, hurting our ability to fine-tune the embedding representation.  
To resolve these issues, we fine-tune the 8-bit version of \SFRmistral~model using QLoRA~\cite{dettmers2023qlora} with a batch size of 24 per GPU. 
Since there is no need for model sharding, we can now use simple data parallel approaches (e.g., ZeRO-1) and are able to achieve linear scaling up to 400 nodes (1,600 GPUs) of the Polaris supercomputer (see \autoref{sec:scaling_results} for details). 

We use contrastive loss to improve our embedding models, which benefits from large batch sizes~\cite{bachman2019amdim}. Naively increasing batch size by reducing loss from all ranks and applying contrastive or LI loss is limited by dense embeddings becoming large with increased world size, affecting LI more acutely than contrastive losses. We apply LI loss to the local rank only due to its memory and computational demands. For contrastive loss, pooled embeddings from all ranks are gathered, and loss is calculated with the local rank compared to $\min(N, W)$ samples, where $N$ is the maximum to consider and $W$ is the total samples across all ranks. The total loss per iteration is $L = (L_{LI} + L_{C}) / 2$, where $L_{LI}$ is maxsim loss~\cite{khattab2020colbert} and $L_{C}$ is contrastive loss~\cite{gao-etal-2021-simcse}. This ``ColTrast'' loss method is illustrated in \autoref{fig:coltrast-workflow}. Following this fine-tuning algorithm produces an embedding representation that clusters related scientific concepts (discussed further in \autoref{sec:eval}).

\begin{table}[ht]
    \centering
    \caption{Composition of the ColTrast finetuning dataset with a further description provided in \autoref{subsubsec:coltrast_trainset}.}
    \resizebox{\columnwidth}{!}{
    \begin{tabular}{|l|c|c|c|}\hline
        Domain & PDF Count & Chunk Count & Generated questions\\\hline
        Peptides      & \num{10124} & \num{123685}   &   \num{70866} \\
        Cancer    & \num{18034} & \num{208803}   &   \num{114535} \\
        COVID-19    & \num{54592} & \num{519412}   &   \num{270493} \\\hline
        Total    & \num{82750} & \num{851900}   &   \num{455894} \\\hline
    \end{tabular}
    }
    \label{tab:pdf_comp}
\end{table}

\subsubsection{ColTrast Finetuning Dataset}
\label{subsubsec:coltrast_trainset}

\begin{figure}[ht]
    \centering
    \includegraphics[width=\columnwidth]{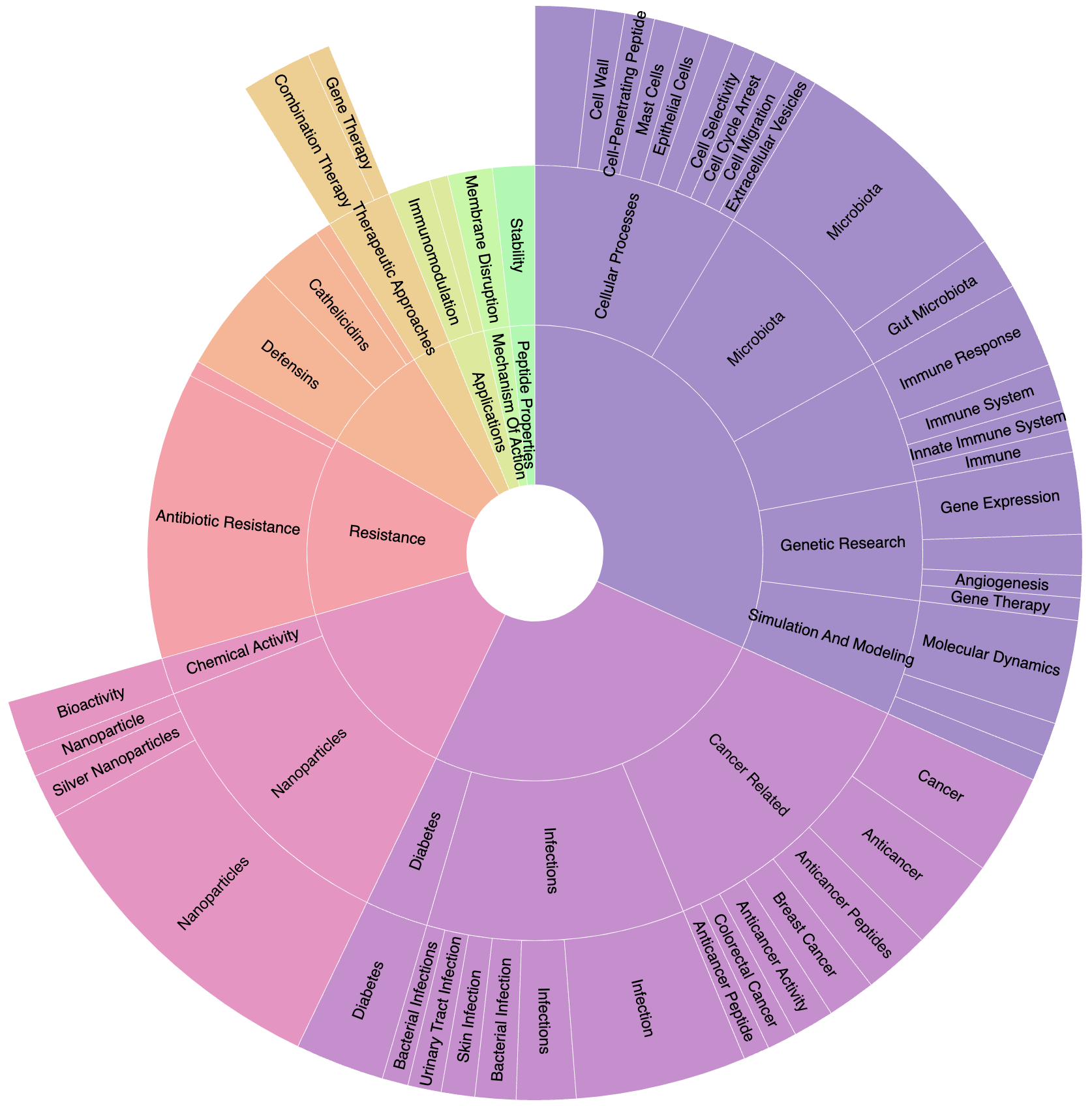}
    \caption{Sub-domain distribution of the antimicrobial peptide dataset for ColTrast finetuning.}
    \label{fig:plcdataset} 
\end{figure}

We finetune ColTrast on three scientific paper datasets, totaling 82,750 papers, that we curated. These papers span diverse topics, including low-dose radiation and cancer biology, antimicrobial peptide literature, and SARS-CoV-2 research. The high-level composition of the combined dataset is summarized in \autoref{tab:pdf_comp}, and the sub-domain distribution of the antimicrobial peptide dataset is illustrated in \autoref{fig:plcdataset} for additional clarity on its diversity.

To enhance the effectiveness of our retrieval system, we apply a semantic chunking algorithm to the text of these papers. Semantic chunking is the process of dividing the content of a paper into concise, coherent segments or "chunks" that each encapsulate a complete semantic idea. This technique involves iterating over the encoded sentences using \SFRmistral~\cite{Meng_SFR_Mistral} and continuously adding to a segment as long as the cosine similarity between consecutive sentences remains above a predetermined threshold. This method not only aids in maintaining the contextual integrity of the information but also optimizes the retrieval process by ensuring that each chunk represents a distinct concept or topic. Such organization of text before vectorization is widely recognized to improve the precision and relevance of retrieved documents, thereby advancing the state of the art in document parsing for LLM-based retrieval systems.

For each semantic chunk created, we then employ Mistral-7B-Instruct-v0.2~\cite{jiangMistral7b2023} to generate a high-level question that uses the chunk as a reference or resource for its answer. This process results in a collection of 455,894 question-chunk pairs, collectively referred to as the ColTrast fine-tuning dataset, as shown in \autoref{tab:pdf_comp}. This structured approach not only improves the efficiency of our retrieval system but also significantly enhances the quality of generated questions, facilitating more accurate and contextually relevant responses.

\subsection{Distributing  HiPerRAG with HPC and Warmstart Optimization}
\label{sec:warmstart}

We use the Parsl parallel programming library~\cite{babuji19parsl} to distribute the execution of our PDF parsing, semantic chunking, chunk encoding, and question sampling workflows. Parsl supports execution across diverse HPC resources and schedulers and has been shown to scale to hundreds of thousands of workers and thousands of tasks per second.
However, standard Parsl assumes that tasks are pure functions, which is not conducive to persisting shared data structures, such as ML models, between tasks.
For example, loading models for semantic chunking is an I/O-heavy operation and takes longer when more nodes read the same weight files concurrently: nearly nine minutes at worst, as shown in \autoref{tab:sem-chunk-load-times}.

Thus, we implement a model registry that makes Parsl workers stateful actors that can persist a shared state across task invocations.  The registry is implemented as a module-level global singleton variable that caches the return result of a Python function or class initialization in a dictionary using a hash of the arguments and keyword arguments as a unique key. The registry permits only one object at a time. Before a new object is created and registered, any existing object is automatically destroyed to free up shared resources, such as GPU memory~\cite{registery}. In our use case, the registry captures a reference to the model upon its first initialization within a worker process (cold start), then returns the cached model for all subsequent task invocations (warm start). This method enhances system utilization by minimizing I/O overheads, amortizing the cost of model loading, and keeping the model in device memory (e.g., CUDA memory) to reduce memory transfers from host to device.

\begin{table}[ht]
\caption{Semantic chunking model load times on Polaris and Sunspot increase at larger scales. The 2- and 128-node values correspond to Polaris, while the 4- and 96-node values correspond to Sunspot. All times in seconds.}
\centering
\resizebox{\columnwidth}{!}{%
\begin{tabular}{|l|c|c|c|c|c|}
\hline
  & \multicolumn{5}{c|}{\textbf{Nodes}} \\
\textbf{System}  & \textbf{2 / 4}   & \textbf{32}     & \textbf{64}     & \textbf{96 / 128} & \textbf{256} \\
\hline
Polaris & 36 $\pm$ 21  & 88 $\pm$ 28  & 114 $\pm$ 24  &  130 $\pm$ 23 & 362 $\pm$ 88 \\
Sunspot & 172 $\pm$ 73 & 218 $\pm$ 107 & 209 $\pm$ 160 & 535 $\pm$ 5   & ---         \\
\hline
\end{tabular}
}
\label{tab:sem-chunk-load-times}
\end{table}

\section{HiPerRAG Evaluation}\label{sec:eval}

It is important to reiterate that all RAG systems are made up of multiple independent components. Accordingly, \aglimmer~consists of the \OREO~ document parser, the ColTrast query-aware encoder, and an arbitrary generator model to produce a response in light of the relevant content retrieved. Our discussion in this section commences with an evaluation of \OREO~ in \autoref{subsec:oreo_eval}. The evaluation of ColTrast in retrieval accuracy will follow in \autoref{subsec:coltrast_eval}. \autoref{sec:qa-datasets} will introduce the scientific Q/A and retrieval accuracy evaluation datasets that this work contributes. \autoref{subsec:scientific-qa-eval-results} will conclude with an evaluation of various ~\aglimmer~ configurations on five scientific Q/A benchmarks, two of which are our contribution.

\subsection{Oreo Document Parsing Evaluation }\label{subsec:oreo_eval} 

We evaluate \OREO, Marker, and Nougat on $n=100$ scientific documents spanning eight domains: mathematics, physics, chemistry, biology, engineering, medicine, economics, and computer science, and six publishers: ArXiv, BioRxiv, MedRxiv, BMC, MDPI, and Nature. Ground truth text is sourced from the full-text HTML versions of the articles.

\begin{table}[ht]
    \caption{Accuracy and throughput of image-based parsers on diverse scientific documents.}
    \centering
    \resizebox{\columnwidth}{!}{%
    \begin{tabular}{|l|c|c|c|}
    \hline
                        & \textbf{Oreo (ours)} & \textbf{Nougat} & \textbf{Marker} \\ \hline
    \textbf{Accuracy (BLEU) [\%]} & 46.34         & 46.42           & \textbf{56.90}           \\ \hline
    \textbf{Accuracy (CAR) [\%]}  & \textbf{73.92}         & 70.92           & 73.51           \\ \hline
    \textbf{Throughput [PDFs/GPU sec]} & \textbf{0.55}          & 0.12            & 0.09            \\ \hline
    \end{tabular}
    }
    \label{tab:accuracy_throughput}
\end{table}

Parsing quality is measured using bilingual evaluation understudy (BLEU) and character accuracy rate (CAR). BLEU evaluates $n$-gram overlap between parsed and ground truth text \cite{papineni2002bleu}. CAR measures character-level precision, particularly valuable for scientific equations and numerical data \cite{zheng2015character}. 

The results in \autoref{tab:accuracy_throughput} show that Marker achieves the highest BLEU score but has the lowest throughput. Nougat offers a competitive BLEU score but the lowest CAR. Notably, \OREO{} excels with the highest CAR and throughput while approximately matching Nougat's BLEU performance. Therefore, \OREO~ maximizes the number of accurately parsed tokens, making it a suitable option for large-scale document parsing.

\subsection{ColTrast Retrieval Evaluation }\label{subsec:coltrast_eval} 

\begin{table}[t!]
    \caption{Model performance on the BioSynthQP dataset and evaluation split of data. Abbreviations used include: BSQP - BioSynthQP dataset; Eval - the evaluation split; M - Salesforce Research/SFR-Mistral base model; B - BERT base model; Q - quantized low-rank approximation (QLoRA) parameter-efficient fine-tuning; ND - No distributed all-gather in contrastive loss; CL - contrastive loss only; LI - Late-Interaction Loss only; S -  Model with ColTrast loss (CL + LI). }
    \centering
    \resizebox{\columnwidth}{!}{%
    \begin{tabular}{|l|c|c|c|c|}
    \hline
                            &  P@10     & MRR@10    & MRR@20     & Top-20 \\
    Name                    &  BSQP      & BSQP      & Eval       & Eval \\
    \hline
    \makecell[l]{ColTrast-M-Q-S} &  0.74  & 0.35  & \textbf{0.93} & 0.62 \\
    \makecell[l]{ColTrast-M-Q-ND-S} & \textbf{0.81} & 0.51 & 0.90 & \textbf{0.98} \\
    \makecell[l]{ColTrast-B-S} &  0.33 & 0.03 & 0.21 & 0.05 \\
    \makecell[l]{ColTrast-B-ND-S}  & 0.72 & 0.46 & 0.84 & 0.96 \\
    \makecell[l]{BERT-CL-S}    &  0.28 & 0.45 & 0.184 & 0.04 \\
    \makecell[l]{BERT-LI-S}    &  0.03 & 0.003 & 0.06 & 0.003 \\\hline
    PubMedBERT       & 0.72 & \textbf{0.65} & 0.22 & 0.39 \\
    ColBERT-v2       & 0.31 & 0.11 & 0.371 & 0.97 \\
    BERT-Base        & 0.31 & 0.03 & 0.21 & 0.37 \\
    \hline
    \end{tabular}
    }
    \label{tab:embedder_ablations}
\end{table}


To evaluate the ColTrast retrieval approach, we utilize two datasets: (1) our BioSynthQP dataset described in \autoref{sec:biosynthqp}, and (2) a 5\% held-out evaluation dataset from the ColTrast training set, as described in \autoref{subsubsec:coltrast_trainset}. Rather than sampling the evaluation set randomly, we ensure that the training and evaluation sets contain passage/query pairs from distinct scientific documents. \autoref{tab:embedder_ablations} shows results of training with different model architectures and loss paradigms against the evaluation datasets.  
When provided with a question, the documents are ranked according to the cosine similarity between the question and document embeddings.  


We report the precision as $P = N_{rel} / N_{ret}$ for the number of relevant retrieved documents, $N_{rel}$, and the total number of retrieved documents, $N_{ret}$.  In the case of P@10, $N_{ret}$ is set to 10.  MRR@10 is calculated as $MRR = N_q^{-1} \sum_i \textrm{rank}_i^{-1}$ where rank$_i$ is the ranking of the positive paired document for the $i^{th}$ query, $N_q$ is the total number of queries, and @10 sets that if the positive document is ranked beyond 10, that term of $MRR$ is zero.  
All models in the upper section of \autoref{tab:embedder_ablations} were trained with the IA$^3$\cite{liu2022ia3} parameter-efficient fine-tuning (PEFT) method, except ColTrast-M-Q-\*, since the QLoRA approach requires LoRA PEFT\cite{hu2022lora}.  We performed ablations to examine the effectiveness of the ColTrast combined loss method. BERT-CL-S uses contrastive loss only, BERT-LI-S only has late-interaction loss, and $<$\textit{Model name}$>$-ND-S denotes models with the ColTrast loss method but excluding the all-gather for the contrastive loss. The base model used for fine tuning is denoted by ``M'' for SFR-Mistral and ``B'' for BERT-base. 
We observe that performance on evaluation metrics is improved with the ColTrast loss method, as opposed to utilizing only LI or contrastive losses and that, for these models where a decent batch size is allowed, ND performs comparably to the distributed loss.  The models in the lower half of the table are pre-existing approaches to the embedding and retrieval task included for comparison.




\subsection{Scientific Question-Answering and Retrieval Accuracy Evaluation Datasets}
\label{sec:qa-datasets}
Question-answering (QA) datasets are pivotal for both training and evaluating LLMs in various scientific domains. These benchmarks are especially valuable for gauging a model's depth of domain-specific knowledge, its ability to understand intricate questions, and its proficiency in quantifying uncertainty during the generation of open-ended responses \cite{khot2018scitail}. In Sections~\ref{sec:protein_function_qa} and \ref{sec:protein_interaction_qa}, we present two novel scientific QA datasets that we used to evaluate RAG configurations within the \aglimmer{} framework. In \autoref{sec:biosynthqp} we present a synthetic retrieval accuracy evaluation dataset, which was used for assessing retrieval accuracy of query-aware scientific encoders, such as ColTrast that we present in this work. 

\subsubsection{ProteinInteractionQA}
\label{sec:protein_interaction_qa}

We compiled a dataset of \num{16009} antimicrobial peptides from three sources: The Antimicrobial Peptide Database (APD) \cite{wang2004apd}, the Database of Antimicrobial Activity and Structure of Peptides (DBAASP) \cite{pirtskhalava2021dbaasp}, and the Database of Antimicrobial Resistance Peptides (DRAMP) \cite{shi2022dramp}. For each peptide, we utilized the UniProt API \cite{uniprot2021uniprot} to retrieve data on proteins that interact with them. This effort identified \num{7591} peptides with known interactants. Utilizing the instruction-tuned Mistral7B model \cite{jiang2023mistral}, we generated multiple-choice questions from this experimentally validated data. The resulting task, ProteinInteractionQA, poses questions such as ``What protein does <peptide-name> interact with?'' The correct answer includes all known interactants from UniProt, while the distractors comprise non-interactants randomly selected from other peptides in the dataset. It is worth noting that formatting our curated data into a Q/A task did not require a state-of-the-art LLM. Mistral7B proved sufficient for this purpose, thanks to its ability to follow instructions and produce structured output effectively.

\subsubsection{ProteinFunctionQA}
\label{sec:protein_function_qa}

To curate the ProteinFunctionQA, we downloaded functional descriptions for \num{17646} antimicrobial peptides from the UniProt database. We then employed the Mistral7B \cite{jiang2023mistral} instruction-tuned LLM to generate multiple-choice questions based on this experimentally annotated data. The correct answer reflected the ground-truth function of the peptide as retrieved from UniProt, while the distractors were incorrect functions sampled from other peptides within the dataset. The task requires selecting the correct function for a given peptide.

\subsubsection{BioSynthQP}
\label{sec:biosynthqp}

We created this synthetic biomedical data set by using GPT-4 prompt engineering with the express purpose of evaluating retrieval accuracy. BioSynthQP features scientific questions surrounding subdomains of medicine such as virology, oncology, and cardiology. For each subdomain, we generate a set of questions, along with 10 paragraphs for each question with a decreasing level of accuracy and relevance. In this context, the most relevant paragraph constitutes an answer that contains keywords, empirical results, and citations. The least relevant paragraph, on the other hand, digresses away from the question, lacks scientific basis, and citations, but still is broadly related to medicine. We apply human supervision to the postprocessing of the synthetic content to guarantee the accuracy of the quality labels assigned to each sample. Consequently, BioSynthQP presents a demanding retrieval task that evaluates an encoder’s capability to generate embeddings that accurately capture quality characteristics within samples from the same domain.

\subsection{Scientific Q/A Evaluations}
\label{subsec:scientific-qa-eval-results}
We analyze the performance of our scientific Q/A from three angles: composition of the retrieval corpus, encoder model employed for retrieval, and the generator that leverages the retrieved content to answer questions. The two scientific corpora that we curated for this analysis differ greatly in terms of their domain composition and document size. As shown in \autoref{tab:pdf_comp}, PLC consists of \num{10124} PubMed articles about proteins. We encode this corpus with PubMedBERT, an encoder-based LLM specifically fine-tuned on PubMed articles \cite{Gu_2021}. On the other hand, Scientific Literature Corpus (SLC) contains over 3.6 million articles across numerous domains. We encode SLC with our Coltrast-B-S and Coltrast-M-Q-S encoders, which have been fine-tuned on a comprehensive scientific dataset (\autoref{subsec:coltrast_eval}). We utilize instruct-finetuned versions of two generators Mistral-7B and Mixtral-8x7B, both of which are instruct-finetuned. \autoref{tab:model_comparison} compares encoder models with respect to accuracy on five scientific Q/A tasks. SciQ \cite{sciq} covers crowd-sourced science exam questions on Biology, Chemistry, and Physics. PubMedQA contains biomedical research questions. LitQA \cite{Lala2023PaperQA} also focuses on the biomedical domain, but features questions that can only be answered from knowledge in full-text papers. ProteinInteractionQA and ProteinFunctionQA are presented in this work (\autoref{sec:qa-datasets}). We also provide baseline results where we do not use retrieval.

We observe that the Mixtral8x7B model which retrieves with PubmedBERT \cite{pubmedbert} outperforms GPT-4 (75.2\% accuracy) on the PubmedQA benchmark \cite{pubmedqa2023} despite having access to a much smaller biomedical corpora (PLC), and having a smaller parameter count (14-billion parameters utilized at inference). Additionally, this model also outperforms PubmedGPT \cite{bolton2022pubmedgpt}, a domain-specific LLM with 2.7 billion parameters pre-trained from scratch on biomedical data. This result suggests that retrieval-augmented generation at scale can improve the accuracy of LLMs in a data-efficient manner and render general-purpose LLMs more performant than domain-specific models on scientific QA tasks. Moreover the Mixtral8x7B model which uses the (ColTrast-B-S) encoder answers 90\% of the questions in the comprehensive SciQ dataset. ColTrast-B-S retrieval and SLC corpora lead to a 12\% improvement of this model on that benchmark. This result is consistent with our hypotheses that combining a novel metric learning-based retrieval strategy with the ability of retrieving from millions of scientific articles can equip LLMs with domain-specific knowledge beyond their training data.

\begin{table}[ht]
\caption{Comparison of different encoder models with Protein Literature Corpus (PLC) and Scientific Literature Corpus (SLC). Results are reported as accuracy (\%) on multiple-choice scientific question-answering benchmarks. Abbreviations used include: PMB - PubMedBERT Encoder; ColTrast-B-S - ColTrast encoder based on BERT-base; ColTrast-M-Q-S - ColTrast encoder based on SFR-Mistral with QLoRA.}
\label{tab:model_comparison}
\centering
\resizebox{\columnwidth}{!}{
\begin{tabular}{|l|c|c|c|c|c|c|}
\hline
\rotatebox{90}{\textbf{Corpus}} & \textbf{Generator (Encoder)} & \rotatebox{90}{\textbf{SciQ}} & \rotatebox{90}{\textbf{PubmedQA}} & \rotatebox{90}{\textbf{LitQA}} & \rotatebox{90}{\textbf{Protein-}} \rotatebox{90}{\textbf{InteractionQA}} &
\rotatebox{90}{\textbf{Protein-}}\rotatebox{90}{\textbf{FunctionQA}} \\
\hline
\multirow{2}{*}{\textbf{PLC}}
& \makecell{Mistral (PMB)} & 0.80 & 0.45 & 0.32 & 0.44 & 0.32 \\
& \makecell{Mixtral8x7B (PMB)} & 0.88 & \textbf{0.76} & 0.34 & 0.70 & 0.52 \\
\hline
\multirow{3}{*}{\textbf{SLC}}
& \makecell{Mistral (ColTrast-B-S)} & 0.82 & 0.45 & 0.32 & 0.40 & 0.32 \\
& \makecell{Mistral (ColTrast-M-Q-S)} & 0.85 & 0.44 & 0.24 & 0.43 & 0.332 \\
& \makecell{Mixtral8x7B (ColTrast-B-S)} & \textbf{0.90} & 0.75 & 0.34 & 0.66 & \textbf{0.526} \\
\hline
\multirow{2}{*}{\textbf{BASE}}
& \makecell{Mistral} & 0.88 & 0.59 & \textbf{0.40} & 0.44 & 0.36 \\
& \makecell{Mixtral8x7B} & 0.78 & 0.73 & 0.38 & \textbf{0.70} & 0.36 \\
\hline
\end{tabular}
}

\end{table}



\input{tables/table-systems}

\section{HiPerRAG Scaling and Performance}\label{sec:scaling_results}

We conduct extensive scaling experiments on document parsing, semantic chunking, and encoder fine-tuning steps of \aglimmer~. In \autoref{sec:hardware_specs} we describe the HPC platforms on which we conducted our experiments. \autoref{sec:5.2_scaling_experiments} will discuss our experiments and their results in detail.

\subsection{Hardware Platforms Utilized}
\label{sec:hardware_specs}


We evaluate the performance and scale on three diverse supercomputing systems: Polaris at the Argonne Leadership Computing Facility (ALCF), Sunspot at ALCF, and Frontier at the Oak Ridge Leadership Computing Facility (OLCF). In the November 2024 Top-500 list \cite{top500nov2024}, Polaris is ranked \#47 with a peak of 34 PFLOPS
and Frontier is ranked \#2 with a peak of 2.055 EFLOPS. \autoref{tab:system-table} compares the three systems used for evaluation.

\textbf{Polaris} is an HPE Apollo Gen10+ system with 560 nodes interconnected by an HPE Slingshot-11 network with a Dragonfly topology.
Each node consists of an AMD “Milan” processor with 32 cores and 512GB of system memory. four 40~GB NVIDIA A100 GPUs, and two Slingshot-11 25~GB/s network adapters. 
Each NVIDIA A100 GPU is capable of achieving a peak of 19.5 TFLOPS in FP32, 156 TFLOPS in TF32, and 312 TFLOPS in FP16 and BF16.


The \textbf{Sunspot} Test and Development System (TDS) 
is a 128 node HPE Cray EX system. Each node consists of two Intel Sapphire Rapids Xeon CPU Max Series and six Intel Ponte Vecchio Data Center GPU Max Series, as has 1~TB memory
Each Xeon CPU has 52 physical cores (two hardware threads per core). GPUs have 128~GB of high-bandwidth memory (HBM2e) memory and are connected within each node by an all-to-all interconnect. Each node has 8$\times$ HPE Slingshot-11 NICs, providing an injection bandwidth of 200~GB/s. 

\textbf{Frontier} is an HPE Cray EX system comprised of 74 Olympus rack, 128-AMD-compute-node, HPE cabinets, for a total of \num{9408} AMD compute nodes.
Each compute node comprises a single 64-core AMD CPU (two hardware threads per core), 512~GB of DDR4 memory, and four AMD MI250X GPUs.
Each GPU has two Graphics Compute Dies (GCDs) for a total of 8 GCDs per node; GCDs have 64~GB of HBM2e. The nodes are connected with 4$\times$ HPE Slingshot-11 25 GB/s NICs providing a node-injection bandwidth of 100~GB/s.



\subsection{Scaling the HiPerRAG Workflow}
\label{sec:5.2_scaling_experiments}

We trained the embedding retriever models with the Huggingface Accelerate library and the Huggingface trainer.  Aside from implementation simplicity, the trainer also supplies estimates of total FLOPs and run time at each checkpoint.  From these estimates, and a known world size, we  estimated TFLOPS per GPU. The trainer (after a complete run) also supplies estimates of training steps per second.  We use these metrics to derive the performance results presented in the next section.

\begin{table}[b]
    \centering
    \caption{Estimated floating point operations and throughput for embedding/retriever model architectures during 100-step scaling runs using 400 GPUs on Polaris supercomputer ($<$\textit{model}$>$-P).  BERT-base-LoRA-F was run on 2400 GCDs (1200 GPU) on Frontier supercomputer. Throughput is in samples per second, total samples (S$_T$) in millions of samples.
    } 
  \resizebox{\columnwidth}{!}{%
        \begin{tabular}{|l|r|r|r|r|}
        \hline
                             &      TFLOPS/    &       FLOPs     & &  \multicolumn{1}{c|}{S$_T$} \\
         Name                     &     GPU    &       (10$^{18}$)    & Samples/s&  (10$^{6}$) \\
       \hline
        Mistral-LoRA-P             &        0.71        &       2.40      &   152.2              &  0.64  \\
        Mistral-QLoRA-P            &        93.80        &       160.00      &   3496.4              &  3.84  \\
        BERT-base-LoRA-P           &        7.95        &       0.67      &   \num{24157.0}               &  2.56  \\
        BERT-base-LoRA-F        &        1.92        &       3.10            &   8654.8              &  7.68\\
        \hline
        \end{tabular}
    }   
    \label{tab:training_flops}
\end{table}




\begin{figure*}[ht]
    \centering
    \includegraphics[width=\textwidth]{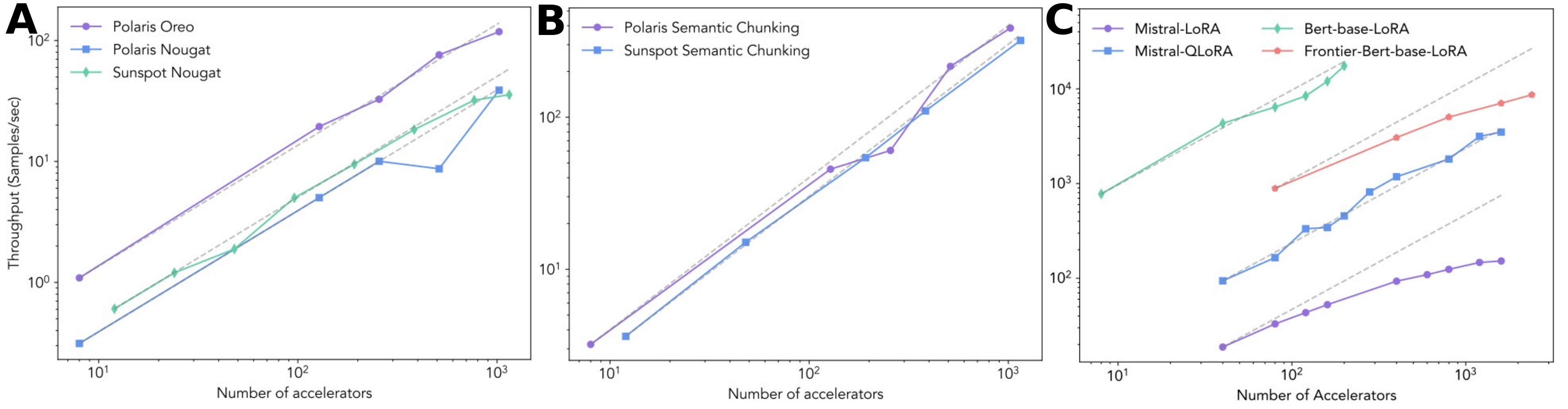}
    \caption{A) Strong scaling results for PDF parsing workflows on Polaris and Sunspot; B) Strong scaling results for semantic chunking workflows on Polaris and Sunspot; C) Strong scaling results for different model architectures on Polaris and Frontier. Unless noted, a run was accomplished on Polaris. Note that one GPU in this figure on Frontier corresponds to one GCD.}
    \label{fig:strong-scaling}
\end{figure*}

\autoref{fig:strong-scaling}-A shows the performance of the Oreo and Nougat parsers as we strong scale on Polaris and Sunspot by increasing the number of accelerators to 1024 GPUs on Polaris (256 nodes) and to 1152 tiles on Sunspot (96 nodes, each with 6 GPUs and 12 tiles). On Polaris at 1024 accelerators, Oreo achieves 118.2 samples/s and Nougat 38.9 samples/s: a 3$\times$ improvement in throughput with Oreo over Nougat.
This improved throughput is key to realizing our vision of being able to parse and extract, in a scalable manner with high quality, all of the scientific literature generated. 

Next, we compare the performance of Nougat processing on Polaris and Sunspot. We observe a throughput of 35.62 samples/s on 1152 tiles (96 nodes) of Sunspot and 38.9 samples/s on 1024 GPUs (256 nodes) of Polaris. At a node-to-node level, we observe a 2.36$\times$ improvement in performance on Sunspot over Polaris. This result is expected as we have more GPUs and memory on each Sunspot node. At an accelerator level, comparing a single A100 GPU to a Intel Max GPU tile, we currently observe a 26\% improvement for an A100 GPU in comparison to a tile on the Intel Max Data Center GPU. We attribute this primarily to the current state of optimization of the software stack here. In all cases, we observe a linear scaling in throughput as we scale out, which we attribute to the embarrassingly parallel nature of the parsing workflows. 

For end-to-end parsing, Oreo achieves 40.7 TFLOPS/GPU peak with a sustained average of 37.2 TFLOPS/GPU on an NVIDIA A100, while Nougat achieves 0.43 TFLOPS/GPU peak and a sustained average of 0.28 TFLOPS/GPU. Thus, Oreo achieves 94.6$\times$ better compute performance than Nougat while sustaining a 4$\times$ improvement in throughput. This result is expected as Oreo follows a compartmentalization strategy that adds compute requirements in two ways. First, layout detection and combination of text items require tensor operations. Second, once text items are transcribed, they need to be mapped to their proper position. These innovations enable Oreo to process efficiently multiple file formats and multi-modal document assets including figures, tables, equations, and code.

We next compare the strong scaling performance of our end-to-end Semantic Chunking workflow on Polaris and Sunspot. As seen from \autoref{fig:strong-scaling}-B, we observe linear scaling, in terms of throughput in samples/s, as we scale with the number of accelerators. This result is expected given the embarrassingly parallel nature of the workflow. We observe a slight dip in performance at 256 accelerators (64 nodes), likely due to storage system limitations when reading the data to be chunked. We observe a throughput of 385.7 samples/s on 1024 accelerators (256 nodes) of Polaris. On Sunspot, we observe a throughput of 319.8 samples/s on 1152 accelerators (96 nodes). Normalizing these at an accelerator level, we observe a 35\% improvement in achievable throughput on an A100 GPU in comparison to a tile of an Intel GPU. In terms of a node-to-node comparison, we achieve a 2.2$\times$ improvement on a Sunspot node in comparison to a Polaris node.


\autoref{fig:strong-scaling}-C shows strong scaling for training encoder models with samples/s as our throughput metric.  Polaris runs include major model architectures used here: BERT base and mistral. Each model is trained using parameter-efficient fine-tuning---for computational efficiency and to help the model retain its knowledge from pretraining. Early experiments showed less than ideal scaling for Mistral-LoRA; since the model is too large for a single GPU, we sharded it across ranks by using Deepspeed ZeRO-3, and further employed parameter and optimizer state offloading in order to maximize batch size.  To overcome this poor scaling, we adopted QLoRA training, where the model is first quantized to reduce memory pressure and then a LoRA adapter is applied as trainable parameters in full (bf16) precision.  The Mistral-QLoRA approach yielded twofold benefits: first, the reduced memory footprint allowed us to load the full Mistral model into a single Nvidia-A100 with a batch size of 24; second, the reduced communication overhead compared to ZeRO-3 also enabled ideal scaling up to 400 nodes of Polaris.  

Despite the success of QLoRA on Polaris, we were unable to employ a similar tactic on Frontier, as the BitsAndBytes quantization library required by transformers is not yet supported on MI250X. Performance in terms of TFLOPS/GPU, total FLOPs and sample throughput are enumerated for the scaling runs in \autoref{tab:training_flops}.  Notably, applying the QLoRA approach to \SFRmistral~fine-tuning increases TFLOPS/GPU by 132$\times$, and also achieves the highest TFOPS/GPU and total FLOPs of all measured models. Further, given the total document chunks of the SLC dataset (16.4 M samples), Mistral with QLoRA could iterate an epoch of data in 1.3 hours. Such throughput enables fine tuning on corpora of hundreds of millions of PDF documents.

\section{Conclusions}\label{conclusions}
We presented \aglimmer~ as a scalable scientific knowledge synthesis framework that adapts existing LLMs with fast and efficient neural retrieval approaches to minimize hallucinations. Our methods enhance traditional RAG models by integrating contrastive learning and late interaction techniques to customize the distance metric based on the input query. This enables more accurate retrieval of relevant information from potentially millions of scientific documents and allows the model to effectively navigate a high-dimensional embedding space. By learning semantic connections among diverse domains, the system approaches state-of-the-art performance in scientific question answering benchmarks in a zero-shot manner (i.e., without the need for fine-tuning the generator). This is a significant advance for scientific literature data, where often, a model trained for a specific purpose may need extensive fine-tuning to be performant on other tasks. 

Our work has implications for how knowledge from scientific literature is encoded and represented using LLMs. We have demonstrated a scalable workflow that enables a ``plug and play'' framework for any LLM of choice while enabling retrieval at scale. Semantic chunking allows us to learn viable contexts within scientific literature, while also pointing to (and potentially learning) new semantic interactions, which provide additional contexts for the models to organize emerging concepts or `trends' in scientific research. We intend to use temporal evolution strategies (i.e., by predicting emerging concepts or research themes at a future date, given current scenarios) similar to approaches in~\cite{Lin_2022,Fortunato_2018}. 

A further implication of our work is how knowledge graphs can be instantiated from these representations. The semantic chunking and retrieval frameworks are extensible, and can use any backend model; however, we believe that the implicit representations learned within the embeddings can be used to connect across concepts and scientific domains with rich annotations (that can be enabled by metric learning). These representations then can be used to build an ontology connecting various scientific concepts and disciplines embodied in recent work~\cite{edge2024localglobalgraphrag}. Moreover, it could be used to propose new hypotheses based on existing knowledge.  

Extensions to \aglimmer~ include the capability to pursue multimodality, especially in handling images, video, and audio along with text, which are particularly valuable for parsing scientific articles rich in tables and figures \cite{chen2022muragmultimodalretrievalaugmentedgenerator, zhao2023retrievingmultimodalinformationaugmented, multimodalragjoshi}. Although Oreo can output multimodal data, the lack of established multimodal benchmarks for scientific literature led us to focus on text retrieval as the primary evaluation method for \aglimmer. Future work will experiment with multimodal LLMs as generators to incorporate images and tables that Oreo can parse from a PDF. We expect this to benefit newly emerging workflows---including agentic implementations~\cite{an2024goldenretrieverhighfidelityagenticretrieval, ravuru2024agenticretrievalaugmentedgenerationtime} that begin to power innovative scientific discovery workflows.

\section*{Acknowledgment}
We thank the Argonne Leadership Computing Facility (ALCF) supported by the DOE under DE-AC02-06CH11357 at Argonne National Laboratory and the Oak Ridge Leadership Computing Facility (OLCF) at the Oak Ridge National Laboratory supported under DE-AC05-00OR22725. This project is supported by the Coalition for Epidemic Preparedness Innovations (CEPI) under the Disease X Program, the Argonne Laboratory Directed Research and Development (LDRD) program and US DOE LUCID: Low-Dose Understanding, Cellular Insights, and Molecular Discoveries.

\bibliographystyle{ACM-Reference-Format}
\bibliography{references}

\end{document}

%% file: tables/table-systems.tex
\begin{table}[tb]
\caption{GPU supercomputing systems used for evaluation.}
\resizebox{\columnwidth}{!}{%
\begin{tabular}{|l|l|l|l|}
\hline
                             & \textbf{Polaris} & \textbf{Sunspot} & \textbf{Frontier}          \\ \hline
November-2024 Top 500\#          & 47                & --                & 2             \\       
System size (nodes)             & 560               &  128             & 9408      \\ 
CPU & AMD Milan               & Intel Sapphire Rapids      &       AMD      \\ 
Sockets/node (total cores) & 1 (32) & 2 (104) & 1 (64) \\ 
Node CPU Memory (TB) & 0.5 & 1.0 & 0.5 \\ \hline
GPU            & NVIDIA A100               &  Intel Data Center              & AMD MI250X     \\ 
            &               &  \ \ \ Max (PVC)              &      \\ 
Number of GPUs per node      & 4                & 6             & 4                    \\ 
GPU Memory (GB)         & 40               & 128           & 128            \\ 
GPU Memory Technology         & HBM2               & HBM2e         &    HBM2e           \\ 
GPU Memory BW (TB/s)         & 1.5               & 2.0               &   2.0     \\ 
Interconnect                 & HPE SS-11 & HPE SS-11  & HPE SS-11\\ \hline
NICs per node                 & 2                & 8               & 4        \\ 
Network GB/s per direction     & 25            & 200           &  100          \\ 
\# Nodes (GPUs) scaled & 450 (1800)       &    96 (576, 1152)   &  96 (384,768)    \\ \hline
\end{tabular}%
}
\label{tab:system-table}
\end{table}